\documentclass[aps,twocolumn,showpacs,prb,superscriptaddress]{revtex4}

\usepackage{dcolumn}
\usepackage{amsmath}
\usepackage{epsf}
\usepackage{graphicx}

\newcommand{\btau}{\mbox{\boldmath$\tau$}}
\newcommand{\bnu}{\mbox{\boldmath$\nu$}}
\newcommand{\bomega}{\mbox{\boldmath$\omega$}}

\begin{document}

\title{Slonczewski windmill with dissipation and asymmetry}

\author{Ya.\ B. Bazaliy}

\affiliation{Department of Physics and Astronomy, University of
South Carolina, Columbia, South Carolina 29208, USA}
\affiliation{Institute of Magnetism, National Academy of Science of
Ukraine, 36-b Vernadsky Boulevard, Kyiv 03142, Ukraine.}

\date{October, 2007}

\begin{abstract}
J. Slonczewski invented spin-transfer effect in layered systems in
1996. Among his first predictions was the regime of ``windmill
motion'' of a perfectly symmetric spin valve where the
magnetizations of the layers rotate in a fixed plane keeping the
angle between them constant. Since ``windmill'' was predicted to
happen in the case of zero magnetic anisotropy, while in most
experimental setups the anisotropy is significant, the phenomenon
was not a subject of much research. However, the behavior of the
magnetically isotropic device is related to the interesting question
of current induced ferromagnetism and is worth more attention. Here
we study the windmill regime in the presence of dissipation,
exchange interaction, and layer asymmetry. It is shown that the
windmill rotation is almost always destroyed by those effects,
except for a single special value of electric current, determined by
the parameters of the device.
\end{abstract}


\maketitle

\begin{figure}
  \includegraphics[width=0.45\textwidth]{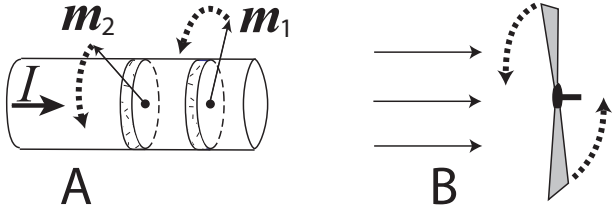}\\
\caption{(A) spin-transfer torques acting on two magnetizations tend
to rotate them in the same direction. Magnetic layers are shaded.
(B) ``windmill'' mechanical analogy.}
  \label{fig:spin_valve}
\end{figure}

Spin-transfer effect as a method of controlling magnetic dynamics by
electric current was suggested by Berger \cite{berger} for domain
wall motion and by Slonczewski \cite{slon96} for spin-valves and
multilayer structures. The unusual property of spin-transfer
interaction found in Ref.~\onlinecite{slon96} was the tendency of
current induced torques to rotate magnetic moments of both spin
valve layers in the same direction, much like the oncoming wind
rotates the wings of a windmill (Fig.~\ref{fig:spin_valve}). If one
assumes that layers have no magnetic anisotropy (crystalline or
shape), are identical, and there is no RKKY exchange or
dipole-dipole interaction between them, the resulting motion is a
perpetual rotation of magnetic moments ${\bf m}_1$ and ${\bf m}_2$
in clockwise or counterclockwise direction, depending on the
direction of electric current $I$ passing through the spin valve. We
will call this type of motion a Slonczewski ``windmill regime''.

Since actual spin-transfer devices have significant magnetic shape
anisotropy, normally the windmill regime is not realized. Instead,
switching between different preferred magnetic configurations was
predicted \cite{slon96} and is intensively studied since then both
experimentally and theoretically. However, the windmill regime still
constitutes an interesting problem due to the following.
Spin-transfer effect can be viewed as reciprocal to the giant
magnetoresistance effect.\cite{baibich88} The resistance of a spin
valve is minimal in the parallel configuration ${\bf m}_1
\uparrow\uparrow {\bf m}_2$. Thus one can hypothesize, that in
response to a current pumped through the valve the magnetizations
will tend to assume this minimal resistance configuration in order
to make electron flow easier. More generally, an idea arises that a
current passing through a metal with paramagnetic impurities will
tend to orient them parallel and create some sort of current-induced
ferromagnetism.\cite{asamitsu1997,masuno2004} The two-magnet device
is the minimal model where the validity of this idea can be tested.
We study the behavior of such a device with arbitrary parameters,
except for the restriction of zero magnetic anisotropy. The results
give a generalized picture of the Slonczewski windmill regime, and
shed some light on the possibility of current-induced
ferromagnetism.

We use the single domain approximation. The magnetic moments ${\bf
m}_i$ ($i = 1,2$) of the layers have time-independent absolute
values $m_i$ and variable directions defined by a unit vector ${\bf
n}_i(t)$. The LLG equations in terms of ${\bf m}_i$ read
\cite{slon96,bjz2004}
\begin{eqnarray}
  \label{eq:LLG_m1}
  \dot{\bf m}_1 &=& \gamma\left(
    {\bf T}_{ex} + {\btau}_1 \right)
    + \frac{\alpha_1}{ m_1} [{\bf m}_1 \times \dot{\bf m}_1] \ ,
 \\
  \label{eq:LLG_m2}
 \dot{\bf m}_2 &=& \gamma\left(
    -{\bf T}_{ex} + {\btau}_2 \right)
    + \frac{\alpha_2}{ m_2} [{\bf m}_2 \times \dot{\bf m}_2] \ ,
\end{eqnarray}
where ${\bf T}_{ex}$ is the exchange torque, $\btau_{1,2}$ are
spin-transfer torques, $\gamma$ is the gyromagnetic ratio, and
$\alpha_{1,2}$ are Gilbert damping constants of the magnets. Note
that in conventional experiments ${\bf m}_2$ is fixed by magnetic
anisotropy, while ${\bf m}_1$ can rotate under the influence of
spin-transfer torque. Magnet number one is then called a ``free
layer'' and magnet number two is called a ``fixed layer'', or spin
polarizer. In the present investigation no restrictions are imposed
on ${\bf m}_2$ and both magnetic moments are treated on equal
footing.

The exchange torque acting on ${\bf m}_1$ is given by ${\bf T}_{ex}
= J [{\bf m}_2 \times {\bf m}_1]$ ($J > 0$ corresponds to
ferromagnetic coupling between the moments). The exchange torque
acting on ${\bf m}_2$ is $- {\bf T}_{ex}$ since we are dealing with
an internal interaction between two moments.

The spin-transfer torques $\btau_{1,2}$ are given by
\begin{eqnarray}
  \label{eq:tau1}
  \dot{\btau}_1 &=&
    u_1 \cdot [{\bf n}_1 \times [{\bf n}_2 \times {\bf n}_1]] \ ,
 \\
 \label{eq:tau2}
 \dot{\btau}_2 &=&
    - u_2 \cdot [{\bf n}_2 \times [{\bf n}_1 \times {\bf n}_2]] \ ,
\end{eqnarray}
with torque strengths
\begin{equation}
u_i = \frac{\hbar}{2}\frac{I}{e}
    g_i[({\bf n}_1 \cdot {\bf n}_2)] \ .
\end{equation}
Here $I$ is the electric current flowing form magnet 2 to magnet 1,
$e$ is the (negative) electron charge, and $g_i[({\bf n}_1 \cdot
{\bf n}_2)]$ are material and device specific spin-polarization
factors. For negligible spin-relaxation in the non-magnetic spacer
between the magnets one has $g_1 = g_2$ (see Ref.
\onlinecite{slon96}). Note that both $u_{1,2}$ are positive when
electrons flow from magnet 2 to magnet 1. The minus sign in front of
the right hand side of Eq.~(\ref{eq:tau2}) reflects the symmetry of
spin-transfer torque.\cite{slon96}

First, we rewrite Eqs. (\ref{eq:LLG_m1}) and (\ref{eq:LLG_m2}) so
that time derivatives are on the left hand side only. Defining ${\bf
T}_1 = {\bf T}_{ex} + \btau_1$, ${\bf T}_2 = -{\bf T}_{ex} +
\btau_2$, we get for $i = 1,2$
\begin{equation}\label{eq:LLG_Phi}
(1+ \alpha_i^2) \dot{\bf m}_i  = \gamma \left( {\bf T}_i +
\frac{\alpha_i}{m_i} [{\bf m_i} \times {\bf T}_i] \right) \ .
\end{equation}
It is convenient to introduce vectors $ \bnu = [{\bf m}_2 \times
{\bf m}_1], \
 {\bf l}_1 = [{\bf m}_1
\times [{\bf m}_2 \times {\bf m}_1]], \
 {\bf l}_2 = [{\bf m}_2 \times
[{\bf m}_1 \times {\bf m}_2]] \ . $ Then
\begin{eqnarray}
 \nonumber
 \dot {\bf m}_1 &=& A_1  \bnu + B_1{\bf l}_1 \ ,
  \\
 \label{eq:dotmi_AB}
 \dot {\bf m}_2 &=& -A_2  \bnu + B_2{\bf l}_2 \ ,
\end{eqnarray}
with
\begin{eqnarray}
 \nonumber
 A_1 &=& \frac{\gamma}{1+\alpha_1^2}
 \left(J - \frac{\alpha_1 u_1}{m_1 m_2} \right) \ ,
 \\
 \nonumber
 A_2 &=& \frac{\gamma}{1+\alpha_2^2}
 \left(J + \frac{\alpha_2 u_2}{m_1 m_2} \right) \ ,
 \\
 \nonumber
 B_1 &=& \frac{\gamma}{(1+\alpha_1^2)m_1}
 \left(\alpha_1 J + \frac{u_1}{m_1 m_2} \right) \ ,
  \\
 \label{eq:AiBi}
 B_2  &=&
  \frac{\gamma}{(1+\alpha_2^2)m_2}
  \left(\alpha_2 J - \frac{u_2}{m_1 m_2} \right) \ .
\end{eqnarray}
Consider now the total magnetic moment of the system ${\bf M} = {\bf
m}_1 + {\bf m}_2$ and calculate the derivative $d{\bf M}^2/dt$.
Using Eq.~(\ref{eq:LLG_Phi}) and the properties $({\bf m}_i \cdot
{\bf l}_i) = 0$, $(({\bf m}_1 + {\bf m}_2) \cdot {\bf l}_i) =
\bnu^2$, we find
\begin{equation}\label{eq:dotM2}
\frac{d {\bf M}^2}{dt} = 2 C [{\bf m}_1 \times {\bf m}_2]^2
\end{equation}
with constant coefficient $C$ that depends on material parameters
and spin-transfer strengths
\begin{eqnarray}\label{eq:C}
 \nonumber
C &=& B_1 + B_2 = \gamma
 \left(\frac{\alpha_1}{(1+\alpha_1^2) m_1}
 + \frac{\alpha_2}{(1+\alpha_2^2) m_2} \right) J
 \\
 &+&
 \frac{\gamma}{m_1 m_2}
 \left(
   \frac{u_1}{(1+\alpha_1^2) m_1} - \frac{u_2}{(1+\alpha_2^2) m_2}
 \right)  \ .
\end{eqnarray}
Since $[{\bf m}_1 \times {\bf m}_2]^2$ is always positive, except in
parallel or antiparallel configurations, we can conclude that after
a transient period the magnetic configuration will reach either the
state of maximal $M$ (i.e., parallel state) for $C
> 0$, or the state of  minimal $M$ (i.e, antiparallel state) for $C
< 0$. Since in both collinear states ${\bf T}_{ex} = 0$ and
$\btau_{1,2} = 0$, the system will come to rest and no ``windmill''
motion will happen. For small spin transfer torques $u_{1,2}$ the
final state will be determined by the sign of $J$ and, as expected,
the device will end up in a configuration corresponding to the
minimum of exchange energy.

The marginal case $C = 0$ is the only situation when the
``windmill'' is possible. According to Eq.~(\ref{eq:dotM2}), the
value of $C$ linearly depends on electric current $I$ through
$u_{1,2}$. The only exception is the singular case when device
parameters satisfy $g_1/[(1+\alpha_1^2)m_1] =
g_2/[(1+\alpha_2^2)m_2]$, and $C$ is current-independent. Thus in
general one can achieve the windmill regime by tuning the current
exactly to the ``marginal'' value $I_w$, such that $C(I_w) = 0$.
Note that this value corresponds to a spin transfer strength of $u_w
\sim \alpha J m_1 m_2$, and since $\alpha \ll 1$ the required spin
torque is much smaller than the exchange torque. The situation is
similar to the switching regime, where spin transfer effect works
against the magnetic anisotropy. In both cases critical values of
spin torque are proportional to the small Gilbert damping
coefficient.

The original discussion of the windmill regime in
Ref.~\onlinecite{slon96} assumed $J = 0$, $\alpha_{1,2} = 0$, and
$m_1 = m_2$. It was found that magnetic moments rotate in the plane
spanned by vectors ${\bf m}_1$ and ${\bf m}_2$ at the initial
moment, and the angle $\theta$ between them remains constant. How
will the windmill motion look in the general situation? At $C = 0$
the total magnetic moment is conserved, ${\bf M}^2 = {\bf m}_1^2 +
{\bf m}_2^2 + 2 ({\bf m}_1 \cdot {\bf m}_2) = {\rm const}$, thus
$\theta$ is constant in general case as well. Since magnitudes of
${\bf m}_1$ and ${\bf m}_2$ are also fixed, constant $\theta$
implies that both vectors will rotate with the same angular
velocity,
\begin{equation}\label{eq:omega_times_mi}
\dot {\bf m}_i = [\bomega \times {\bf m}_i]
\end{equation}
(cf. the theorem on the motion of a rigid body with a fixed point).
To find $\bomega$ we expand it in the basis of vectors $(\bnu,{\bf
m}_1, {\bf m}_2)$ as $\bomega = a\bnu + b_1 {\bf m}_1 + b_2 {\bf
m}_2$ with unknown coefficients $a$ and $b_{1,2}$. Substituting this
form of $\bomega$ into Eqs.~(\ref{eq:omega_times_mi}), using
expressions (\ref{eq:dotmi_AB}) for $\dot{\bf m}_i$, the fact that
for the marginal value of current one has $B_1 = - B_2 \equiv B_w$,
and properties $[\bnu \times {\bf m}_1] = - {\bf l}_1$, $[\bnu
\times {\bf m}_2] = {\bf l}_2$, we find $a = -B_w$, $b_1 = A_2$, and
$b_2 = A_1$
\begin{eqnarray}\label{eq:omega}
\bomega = B_w [{\bf m}_1 \times {\bf m}_2] + A_2 {\bf m}_1 + A_1
{\bf m}_2  \ .
\end{eqnarray}
Since $\dot \bomega = [\bomega \times \bomega] = 0$, $\bomega$ is an
invariant of motion, determined by the initial conditions.

Since $\alpha_{1,2} \ll 1$ and $u_w \sim \alpha J$ at the marginal
point, we can make approximations in expressions (\ref{eq:AiBi}) and
use $A_1 \approx A_2 \approx \gamma J$, $B_1 \approx \gamma(\alpha_1
J m_1 m_2 + u_1)/m_1^2 m_2$, $B_2 \approx \gamma (\alpha_2 J m_1 m_2
- u_2)/m_2^2 m_1$. Equation $C = B_{1} + B_{2} = 0$ then gives the
following values at the marginal point
\begin{eqnarray}
  \label{eq:approxIw}
I_{w} &\approx& \frac{2e}{\hbar}
 \left(\frac{m_1 \alpha_2 + m_2
 \alpha_1}{m_1 g_2 - m_2 g_1} \right)
 J m_1 m_2 \ ,
 \\
 \nonumber
B_{w} &\approx& \gamma\frac{\alpha_1 g_2 + \alpha_2 g_1}{m_1 g_2 -
m_2 g_1} J \ .
\end{eqnarray}
Note that approximation $u_w \sim \alpha J$ is violated when
parameters are close to the degenerate situation $m_1 g_2 - m_2 g_1
= 0$. This is the situation when $C$ is independent of the current
and the windmill regime cannot be achieved. Far away from the
degenerate situation one has
\begin{equation}\label{eq:approximate_omega}
\bomega \approx \gamma J \left(
 \frac{\alpha_1 g_2 + \alpha_2 g_1}{m_1 g_2 - m_2 g_1}
 [{\bf m}_1 \times {\bf m}_2]
 + ({\bf m}_1 + {\bf m}_2)
 \right) \ .
\end{equation}
The first term in parentheses is smaller than the second one by a
factor of $\alpha \ll 1$. Thus vectors ${\bf m}_i$ precess
approximately around the total magnetic moment ${\bf M} = {\bf m}_1
+ {\bf m}_2$. In the presence of damping and spin-transfer $\bf M$
is not conserved and performs a small angle rotation around the
constant vector $\bomega$. In this respect the motion is very
different from Slonczewski's situation at $J = 0$, where $\bf M$ was
performing $360^{\rm o}$ rotations around $[{\bf m}_1 \times {\bf
m}_2]$. The $J = 0$ rotations, however, require the fulfillment of a
condition $g_1/[(1+\alpha_1^2)m_1] = g_2/[(1+\alpha_2^2)m_2]$, which
is, in particular, satisfied in a completely symmetric valve
considered in Ref.~\onlinecite{slon96}.

Finally, we return to Eq.~(\ref{eq:dotM2}) and investigate the $C
\neq 0$ case. It is convenient to rewrite (\ref{eq:dotM2}) in terms
of $x = \cos\theta$
\begin{equation*}
\dot x = C m_1 m_2 (1 - x^2) \ .
\end{equation*}
The solution reads
\begin{equation*}
x(t) = \cos\theta(t) =
 \tanh\left( \frac{t + t_0}{{\rm sgn}[C] \ T_*}  \right) \ ,
\end{equation*}
with
\begin{equation} \label{eq:T*}
T_* = \frac{1}{|C(I)| m_1 m_2} \ ,
\end{equation}
and parameter $t_0$ determined by the initial angle, $\cos\theta_0 =
\tanh(t_0/T_*)$. We conclude that as $t\to \infty$ the system
approaches a collinear configuration with a current dependent
characteristic time $T_*(I)$. The latter diverges in the vicinity of
the marginal current $I_w$.

In conclusion, we studied the motion of a two layer spin-transfer
device with zero magnetic anisotropy. We show that in the presence
of damping, layer asymmetry, and exchange interaction between the
layers the windmill rotation decays with characteristic time
constant $T_*$. The decay time depends on the current pumped through
the device and diverges at a ``marginal'' current $I_w$. For $I \neq
I_w$ the system reaches either a parallel or an antiparallel state
after a transient period. Exactly at the marginal point $I = I_w$
the system performs a perpetual generalized windmill motion.

Interestingly, precession motion analogous to the windmill regime
was also found in multilayers and bilayers with magnetic
anisotropy.\cite{grollier2006,boj2007} In those systems it exists
not at a singular point, but in the whole range of current values.
Thus, rather unexpectedly, anisotropy can be advantageous for the
windmill regime.

Finally, coming to the discussion of the current induced
ferromagnetism, we see that in a two magnet device current can
induce both ferromagnetic and antiferromagnetic order. However, the
situation with only two magnets can be special, and it is necessary
to consider devices with three and more magnets to predict what
happens in the system of many isotropic paramagnetic impurities
under the influence of spin-transfer torques.

\end{document}